\documentclass[intlimits,twoside,a4paper]{article}

\usepackage{amsmath,amssymb}
\usepackage{graphicx}
\usepackage{booktabs}
\usepackage{wrapfig}

\usepackage[T2A]{fontenc}
\usepackage[cp1251]{inputenc}

\usepackage{cmpj2}
\issue{2012}{15}{1}{13703}

\doinumber{10.5488/CMP.15.13703}

\title[Investigations of EPR parameters for AgX:Fe$^{3+}$]{Investigations of the EPR parameters for the tetrahedral
[FeX$_4$]$^{-}$ clusters in  AgX (X${}={}$Cl, Br)}

\author[B.-T. Song \textsl{et al.}]{B.-T. Song\refaddr{ad1}\thanks{E-mail: btsong1@gmail.com}\,, S.-Y.~Wu\refaddr{ad1,ad2}, M.-Q.~Kuang\refaddr{ad1}, Z.-H.~Zhang\refaddr{ad1}}

\authorcopyright{B.-T. Song, S.-Y.~Wu, M.-Q.~Kuang, Z.-H.~Zhang, 2012}

\addresses{\addr{ad1} Department of Applied Physics, University of Electronic Science and Technology of China, \\ Chengdu 610054,
P.R. China%
\addr{ad2} International Centre for Materials Physics,
Chinese Academy of Science, Shenyang 110016, P.R. China}

\date{Received October 12, 2011, in final form January 12, 2012}
\begin{document}

\maketitle

\begin{abstract}
The EPR parameters $g$ factors and the superhyperfine
parameters for the tetrahedral [FeX$_4$]$^{-}$ clusters in
AgX (X${}={}$Cl, Br) are theoretically investigated from the
perturbation formulas of these parameters for a $3d^5$ ion
under tetrahedra by considering both the crystal-field and charge
transfer contributions. The related model parameters are
quantitatively determined from the cluster approach in a uniform
way. The $g$-shift ${\triangle}g$ ($=g-g_{\mathrm{s}}\,$, where
$g_{\mathrm{s}}\approx 2.0023$ is the spin only value) from the charge
transfer contribution is opposite (positive) in sign and much larger
in magnitude as compared with that from the crystal-field one. The
importance of the charge transfer contribution increases rapidly
with increasing the covalency and the spin-orbit coupling
coefficient of the ligand and thus exhibits the order of AgCl${}<{}$AgBr.
The unpaired spin densities of the halogen $ns$,
$np\sigma$  and $np\pi$ orbitals are quantitatively determined from
the related molecular orbital coefficients based on the cluster approach.

\pacs 71.70.Ch, 74.25.Nf, 74.72.Bk
\end{abstract}

\section{Introduction}

Silver halides ({AgX}, with {X=Cl} and
{Br}) containing iron (e.g., {Fe}$^{3+}$) have
interesting electrochemical~\cite{1,2}, magnetic~\cite{3,4}, photocatalytic~\cite{5} and structural~\cite{6} properties and attract extensive attention of
researchers. It is well known that these properties are closely
related to the electronic states and local structures of the
impurity ions in the hosts, which may be studied by means of
electron paramagnetic resonance (EPR) technique. On the other hand,
{Fe}$^{3+}$ among the transition-metal ions belongs to the
half-filled $3d^{5}$ configuration and exhibits
$^{6}{A}_{1{g}}$ ground state with high spin ($S$=5/2) and
quenched orbital angular momentum ($l$ = 0) under weak or intermediate crystal-fields~\cite{7,8}. Therefore, the EPR studies on
{Fe}$^{3+}$ in crystals are of particular importance. For
example, the EPR investigations have been performed for
{Fe}$^{3+}$ doped {AgX}, and the isotropic $g$ factors
and the superhyperfine parameters were also measured decades ago~\cite{9}. Until now, however, the above experimental results have not
been satisfactorily interpreted, except (i) that the centers were
ascribed to the impurity Fe$^{3+}$ on the interstitial sites
in AgX with four nearest neighbour silver vacancies
($V_{\mathrm{Ag}}$) as charge compensation and (ii) that the
superhyperfine parameters were only qualitatively estimated by
fitting the adjustable unpaired spin densities in the previous work~\cite{9}.

As for the Fe$^{3+}$ centers in AgX, the
systems have strong covalency due to the high valence state of the
impurity and the strong covalency of the hosts, which may bring
about significant effect on the EPR parameters (e.g., the $g$
factors and the superhyperfine parameters), as mentioned in some
references for transition-metal impurities in covalent hosts~\cite{10,11}. Thus, further systematic theoretical analysis for the EPR
parameters of AgX:Fe$^{3+}$ are of great
scientific significance. According to the previous studies~\cite{12,13,14},
the ligand orbital (or covalency) and spin-orbit coupling
contributions should be considered for Fe$^{3+}$ (or other
transition-metal ions with high valence states) in the systems with
strong covalency in view of the strong covalency and ligand
spin-orbit coupling interactions (particularly for the ligand
{Br}$^{-}$). Importantly, not only the crystal-field mechanism
related to the antibonding orbitals but also the charge transfer
mechanism related to the bonding (and non-bonding) orbitals can
affect the $g$ factors for these systems~\cite{15,16}. Moreover, the
energy levels of the charge transfer bands may decline with
increasing atomic number of the ligand in the same group of periodic
table~\cite{17}. Thus, the importance of the charge transfer
contributions to the $g$ factor is expected to increase rapidly from
{Cl}$^{-}$ to {Br}$^{-}$ for the same central ion
Fe$^{3+}$ in the studied AgX:Fe$^{3+}$. In order to
clarify the importance of the charge transfer contributions and to
study the EPR spectra of AgX:Fe$^{3+}$ more in detail, the perturbation formula of the $g$ factor is adopted in
this work for a $3d^5$ ion under tetrahedra containing both
the crystal-field and charge transfer contributions based on the
cluster approach. Meanwhile, the superhyperfine parameters of the
ligands Cl$^{-}$ and Br$^{-}$ are also theoretically
studied in a uniform way, with the related unpaired spin densities
determined quantitatively from the cluster approach.

\section{Theory and calculations}

AgX has the NaCl structure. When a
Fe$^{3+}$ ion is doped into the lattice of AgX, it
may occupy the octahedral interstitial site~\cite{9}. This interstital
site in AgX has four nearest neighbour silver ions at the
corners of a cube, and the halogen ligands form a regular
tetrahedron. When impurity Fe$^{3+}$ enters the interstitial
site, the original four nearest neighbour Ag$^{+}$ may escape
to infinity and leave four vacancies due to charge compensation. So,
there are only four nearest neighbour halogen ligands, i.e., the
tetrahedral [FeX$_4$]$^{-}$ clusters (see figure~1 of~\cite{9}).

\subsection*{2.1 Calculations for the $g$ factor}

A $3d^5$ ion under an ideal octahedron may exhibit
the orbital non-degenerate $^{6}{A}_{1{g}}$ ground state
of high spin $S=5/2$~\cite{10,11}. According to extensive studies on
$3d^5$ ions in crystals, the combination of a spin-orbit
coupling and orbital angular momentum interactions may be regarded
as the dominant origin of $g$-shift  $\Delta$$g$ and zero-field
splittings~\cite{18,19}. Applying the Macfarlane's perturbation-loop
method~\cite{20}, the perturbation formula of the $g$-shift containing
both the crystal-field and charge transfer contributions for a
$3d^5$ ion in tetrahedra can be expressed as:
\begin{eqnarray}
\label{eq:1}
&&\Delta g=g-g_{\mathrm{s}}=\Delta g_\mathrm{CF}+\Delta
g_\mathrm{CT}\,,\nonumber\\
&&\Delta g_\mathrm{CF}=-5\zeta_\mathrm{CF}^{\prime2}\left(1/E_{1}^{2}+1/E_{3}^2\right)/6
-\zeta_\mathrm{CF}^{2}/E_{2}^{2}+4\zeta_\mathrm{CF}^{\prime}
\zeta_\mathrm{CF}\left(1/E_{1}+1/E_{3}\right)/\left(5E_{2}\right),\nonumber\\
&&\Delta
g_\mathrm{CT}=(4/5)\left(k_\mathrm{CT}\zeta_\mathrm{CT}/E_\mathrm{n}-k_\mathrm{CT}^{\prime}
\zeta_\mathrm{CT}^{\prime}/E_\mathrm{a}\right).
\end{eqnarray}
Here $E_{1}$, $E_{2}$ and $E_{3}$ are respectively, the energy
differences between the ground $^{6}{A}_{1{g}}$ and the
crystal-field excited
$^{4}{T}_{1}[{t}_{2}^{4}(^{3}{T}_{1})e]$,
$^{4}{T}_{1}[{t}_{2}^{3}(^{2}{T}_{2}){e}^{2}(^{3}{A}_{2})]$
and
$^{4}{T}_{1}[{t}_{2}^{2}(^{3}{T}_{1}){e}^{3}]$
states. They can be expressed in terms of the cubic field parameter
$Dq$ and the Racah parameters $B$ and $C$ for the $3d^5$
ion in crystals: $E_{1}\approx 10B + 6C-10Dq$, $E_{2}\approx 19B +
7C$ and $E_{3}\approx 10B + 6C + 10Dq$. $E_\mathrm{n}$ and $E_\mathrm{a}$ are
those between the ground $^{6}{A}_{1{g}}$ and the charge
transfer excited $^{6}{T}_{1}^{\mathrm{n}}$ and
$^{6}{T}_{1}^{\mathrm{a}}$ states, which are obtained from the
empirical relationships $E_\mathrm{n}\approx 30000 [ \chi(L)- \chi(M)] +
56B/3- 10 Dq$ and $E_\mathrm{a}\approx E_\mathrm{n}-10Dq$~\cite{17}. Here $ \chi(L)
 $ and $\chi (M) $ are, respectively, the optical electronegativities of
the ligand and metal ions.

The subscripts CF and CT denote the related terms in the
crystal-field and charge transfer mechanisms, with the corresponding
spin-orbit coupling coefficients  $\zeta_\mathrm{CF}$, $\zeta_\mathrm{CF}^{\prime}$,
$\zeta_\mathrm{CT}$, $\zeta_\mathrm{CT}^{\prime}$ and the orbital reduction factors
$k_\mathrm{CT}$, $k_\mathrm{CT}^{\prime}$. In view of the charge transfer contribution
to the $g$ factor, one can write the many electron wave functions of
the charge transfer configurations in terms of the eleven-electron
wave functions out of ${t}_{2}^{\mathrm{a}}$,
${t}_{2}^{\mathrm{b}}$ and ${e}^{\mathrm{n}}$, where the
superscripts a, b and n denote the anti-bonding orbitals
(corresponding to the crystal-field mechanism), bonding orbitals
(corresponding to the charge transfer mechanism) and non-bonding
orbitals, respectively. Thus, the ground state
$^{6}{A}_{1{g}}$ is expressed as follows~\cite{11}:
\begin{eqnarray}
\label{eq:2}
&&|^{6}{A}_{1}\frac{5}{2}a_{1}\rangle
=\left[\xi^{+}\eta^{+}\zeta^{+}\theta^{+}\varepsilon^{+}
|\xi^{+}\xi^{-}\eta^{+}\eta^{-}\zeta^{+}\zeta^{-}\right].
\end{eqnarray}

 In the square bracket on the right side of equation~(\ref{eq:2}),
the letters ($\xi $, $\eta$, $\zeta $ and $\theta$, $\varepsilon$) on the left column are ${t}_{2}^{\mathrm{a}}$ and
${e}^{\mathrm{n}}$ orbitals and those ($\xi$, $\eta$ and
$\zeta$) on the right column are ${t}_{2}^{\mathrm{b}}$ orbitals.
There are only two excited configurations
$({t}_{2}^{\mathrm{a}})^{3}({e}^{\mathrm{n}})^{3}({t}_{2}^{\mathrm{b}})^{5}$
(or $^{6}{T}_{1}^{\mathrm{n}}$) and
$({t}_{2}^{\mathrm{a}})^{4}({e}^{\mathrm{n}})^{2}({t}_{2}^{\mathrm{b}})^{5}$
(or $^{6}{T}_{1}^{\mathrm{n}}$) having the non-zero spin-orbit
coupling interactions with the ground state
$^{6}{A}_{1{g}}$. Thus, the $z$ components of
$^{6}{T}_{1}^{\mathrm{n}}$ and $^{6}{T}_{1}^{\mathrm{a}}$
states with the highest $M_{S}$ = 5/2 can be given as~\cite{11}
\begin{eqnarray}
&&|^{6}{T}_{1}^{\mathrm{n}}\frac{5}{2}z\rangle
=\left[\xi^{+}\eta^{+}\zeta^{+}\theta^{+}\varepsilon^{+}
\varepsilon^{-}|\xi^{+}\xi^{-}\eta^{+}\eta^{-}\zeta^{+}\right],\nonumber\\
&&|^{6}{T}_{1}^{\mathrm{a}}\frac{5}{2}z\rangle=-\frac{1}{\sqrt{2}}
\left\{\left[\xi^{+}\xi^{-}\eta^{+}\zeta^{+}\theta^{+}\varepsilon^{+}|\xi^{+}\xi^{-}\eta^{+}\zeta^{+}\zeta^{-}\right]
+\left[\xi^{+}\eta^{+}\eta^{-}\zeta^{+}\theta^{+}
\varepsilon^{+}|\xi^{+}\eta^{+}\eta^{-}\zeta^{+}\zeta^{-}\right]\right\}.
\label{eq:3}
\end{eqnarray}

From the cluster approach, the one-electron basis functions
for a tetrahedral $3d^5$ cluster may be expressed as:
\begin{eqnarray}
&&\Psi_{t}^{x}=N_{t}^{x}\left(\phi_{dt}-\lambda_{\sigma}^{x}\chi_{p\sigma}-\lambda_{\mathrm{s}}^{x}\chi_{\mathrm{s}}-
\lambda_{\pi}^{x}\chi_{p\pi t}\right),\nonumber\\
&&\Psi_{e}^{x}=N_{e}^{x}\left(\phi_{de}-\sqrt{3}\lambda_{\pi}^{x}\chi_{p\pi
e}\right).
\label{eq:4}
\end{eqnarray}
Here, the superscript $x$ (= a or b) denotes the antibonding or
bonding orbitals.  $\phi_{dt}$  and  $\phi_{de}$ are the $d$ orbitals
of the $3d^5$ ion, and $\chi_{p\pi t}$, $\chi_{p\pi e}$,
$\chi_{p\sigma}$ and $\chi_{{s}}$ are the $p$ and $s$ orbitals of ligands.
$N_{t}^{x}$ and $N_{e}^{x}$ are the normalization factors, and
$\lambda_{\sigma}$ and $\lambda_{\pi}$ (or $\lambda_{{s}}$) are the
orbital admixture coefficients. From equation~(\ref{eq:4}), one can obtain the
normalization conditions.
\begin{eqnarray}
&&(N_{t}^{x})^{2}\left[1+(\lambda_{\sigma}^{x})^{2}+({\lambda_{\pi}^{x}})^{2}
-2\lambda_{\sigma}^{x}S_{\sigma}
-2\lambda_{{s}}^{x}S_{{s}}-2\lambda_{\pi}^{x}S_{\pi}\right]=1,\nonumber\\
&&(N_{e}^{x})^{2}\left[1+3(\lambda_{\pi}^{x})^{2}+6\lambda_{\pi}^{x}S_{\pi}\right]=1,
\label{eq:5}
\end{eqnarray}
and the orthogonality relationships
\begin{eqnarray}
&&1+3\lambda_{\pi}^\mathrm{a}\lambda_{\pi}^\mathrm{b}
-3(\lambda_{\pi}^\mathrm{a}+\lambda_{\pi}^\mathrm{b})S_{\pi}=0,\nonumber\\
&&1+\lambda_{\pi}^\mathrm{a}\lambda_{\pi}^\mathrm{b}
+\lambda_{\sigma}^\mathrm{a}\lambda_{\sigma}^\mathrm{b}
+\lambda_{{s}}^\mathrm{a}\lambda_{{s}}^\mathrm{b}
-(\lambda_{\pi}^\mathrm{a}+\lambda_{\pi}^\mathrm{b})S_{\pi}
-(\lambda_{\sigma}^\mathrm{a}+\lambda_{\sigma}^\mathrm{b})S_{\sigma}
-(\lambda_{{s}}^\mathrm{a}+\lambda_{{s}}^\mathrm{b})S_{{s}}=0,\nonumber\\
&&\lambda_{\pi}^\mathrm{a}\lambda_{\pi}^\mathrm{b}
+\lambda_{{s}}^\mathrm{a}\lambda_{{s}}^\mathrm{b}=0.
\label{eq:6}
\end{eqnarray}

Meanwhile, the following approximation relationships are satisfied
for the antibonding orbitals~\cite{19}:
\begin{eqnarray}
N^{2}&\approx&\left[1+6\lambda_{\pi}^\mathrm{a}S_{\pi}+9(\lambda_{\pi}^\mathrm{a})^{2}
S_{\pi}^{2}\right]
\left[1+3(\lambda_{\pi}^\mathrm{a})^{2}+6\lambda_{\pi}^\mathrm{a}S_{\pi}\right]^{-2},\nonumber\\
N^{2}&\approx&\left[1+2\lambda_{\pi}^\mathrm{a}S_{\pi}+2\lambda_{\sigma}^\mathrm{a}S_{\sigma}
+2\lambda_{{s}}^\mathrm{a}S_{{s}}+2\lambda_{\pi}^\mathrm{a}S_{\pi}\lambda_{\sigma}^\mathrm{a}S_{\sigma}
+2\lambda_{\pi}^\mathrm{a}S_{\pi}\lambda_{{s}}^\mathrm{a}S_{{s}}+(\lambda_{\pi}^\mathrm{a})^{2}S_{\pi}^{2}
+(\lambda_{\sigma}^\mathrm{a})^{2}S_{\sigma}^{2}+(\lambda_{{s}}^\mathrm{a})^{2}S_{{s}}^{2}\right]
\nonumber\\
&\times&\left[1+(\lambda_{\pi}^\mathrm{a})^{2}+(\lambda_{\sigma}^\mathrm{a})^{2}
+(\lambda_{{s}}^\mathrm{a})^{2}+2\lambda_{\pi}^\mathrm{a}S_{\pi}
+2\lambda_{\sigma}^\mathrm{a}S_{\sigma}+2\lambda_{{s}}^\mathrm{a}S_{{s}}\right]^{-2}.
\label{eq:7}
\end{eqnarray}
Here $S_{\pi}$, $S_{\sigma}$  and $S_{{s}}$ are the group
overlap integrals between the $d$ orbitals of the $3d^5$ ion
and $p$ or $s$ orbitals of the ligands. $N$ is the average covalency
factor characteristic of the covalency (or orbital admixtures)
between the impurity and ligand ions. In general, the orbital
admixture coefficients increase with an increase of the group overlap
integrals, and one can approximately adopt the proportional
relationship
$\lambda_{\sigma}^\mathrm{a}/S_{\sigma}\approx\lambda_{{s}}^\mathrm{a}/S_{{s}}$ for
the orbital admixture coefficients and the related group overlap
integrals within the same $\sigma$ component.

From the cluster approach, the corresponding spin-orbit
coupling coefficients for the crystal-field mechanism in
equation~(\ref{eq:1}) are expressed as~\cite{21}:
\begin{eqnarray}
&&\zeta_\mathrm{CF}=(N_{t}^\mathrm{a})^{2}\left\{{\zeta_{d}^{0}
+\left[\sqrt{2}\lambda_{\pi}^\mathrm{a}\lambda_{\sigma}^{\mathrm{a}}
-(\lambda_{\pi}^\mathrm{a})^{2}/2-\sqrt{2}A\lambda_{\pi}^\mathrm{a}\lambda_{{s}}^\mathrm{a}\right]
\zeta_{P}^{0}}\right\},\nonumber\\
&&\zeta_\mathrm{CF}^{\prime}=N_{t}^\mathrm{a}N_{e}^\mathrm{a}\left\{\zeta_{d}^{0}
+\left[\lambda_{\pi}^\mathrm{a}\lambda_{\sigma}^\mathrm{a}/\sqrt{2}
+(\lambda_{\pi}^\mathrm{a})^{2}/2
-A\lambda_{\pi}^\mathrm{a}\lambda_{{s}}^\mathrm{a}/\sqrt{2}\right]\zeta_{p}^{0}\right\}.
\label{eq:8}
\end{eqnarray}
Here, $\zeta_{d}^{0}$ and $\zeta_{p}^{0}$ are the spin-orbit
coupling coefficients for a free $3d^5$ and ligand ions,
respectively. $A$ denotes the integral $R\langle
ns|\frac{\partial}{\partial y}|np_{y}\rangle$, where $R$ stands for
the impurity-ligand distance. Similarly, the spin-orbit coupling
coefficients and the orbital reduction factors for the charge
transfer mechanism are written as:
\begin{eqnarray}
&&\zeta_\mathrm{CT}=N_{t}^\mathrm{b}N_{t}^\mathrm{a}\left[\zeta_{d}^{0}
+\left(\frac{\lambda_{\pi}^\mathrm{a}\lambda_{\sigma}^\mathrm{b}
+\lambda_{\pi}^\mathrm{b}\lambda_{\sigma}^\mathrm{a}}{\sqrt{2}}
-\frac{\lambda_{\pi}^\mathrm{a}\lambda_{\pi}^\mathrm{b}}{2}
-{\sqrt{2}}{A\lambda_{{s}}^\mathrm{a}\lambda_{\pi}^\mathrm{b}}\right)\zeta_{p}^{0}\right],\nonumber\\
&&\zeta_\mathrm{CT}^{\prime}=N_{e}^\mathrm{a}N_{t}^\mathrm{b}\left[\zeta_{d}^{0}
+\left(\frac{\lambda_{\pi}^\mathrm{a}\lambda_{\sigma}^\mathrm{b}}{\sqrt{2}}
+\frac{\lambda_{\pi}^\mathrm{a}\lambda_{\pi}^\mathrm{b}}{2}
-\frac{A\lambda_{{s}}^\mathrm{a}\lambda_{\pi}^\mathrm{b}}{\sqrt{2}}\right)\zeta_{p}^{0}\right],\nonumber\\
&&k_\mathrm{CT}=N_{t}^\mathrm{a}N_{t}^\mathrm{b}\left[-\frac{\lambda_{\pi}^\mathrm{a}\lambda_{\pi}^\mathrm{b}}{2}
+\frac{\lambda_{\pi}^\mathrm{a}\lambda_{\sigma}^\mathrm{b}+\lambda_{\pi}^\mathrm{b}\lambda_{\sigma}^\mathrm{a}}{\sqrt{2}}+(\lambda_{\sigma}^\mathrm{a}
+\lambda_{\sigma}^\mathrm{b})S_{\sigma}+(\lambda_{\sigma}^\mathrm{a}+\lambda_{\sigma}^\mathrm{b})S_{\sigma}+(\lambda_{\pi}^\mathrm{a}+\lambda_{\pi}^\mathrm{b})S_{\pi}
-\sqrt{2}A\lambda_{{s}}^\mathrm{a}\lambda_{\pi}^\mathrm{b}\right],\nonumber\\
&&k_\mathrm{CT}^{\prime}=N_{t}^\mathrm{b}N_{e}^\mathrm{a}\left[1+\left(\frac{\lambda_{\pi}^\mathrm{a}\lambda_{\pi}^\mathrm{b}}{2}
+\frac{\lambda_{\pi}^\mathrm{a}\lambda_{\sigma}^\mathrm{b}}{\sqrt{2}}\right)+3\lambda_{\pi}^\mathrm{a}S_{\pi}+\lambda_{\pi}^\mathrm{b}{S}_{\pi}
+\lambda_{\sigma}^\mathrm{b}S_{\sigma}+\lambda_{s}^{\mathrm{b}}S_{{s}}
-\frac{A\lambda_{{s}}^\mathrm{a}\lambda_{\pi}^\mathrm{b}}{\sqrt{2}}\right].
\label{eq:9}
\end{eqnarray}

For the studied [{FeX}$_{4}$]$^{-}$ clusters, the
impurity-ligand distances are $R\approx 2.403$ and 2.500~{\AA}~\cite{22}
for the interstitial sites in {AgCl} and {AgBr},
respectively. Thus, the group overlap integrals and the integral $A$
can be calculated using the distances $R$ and the Slater-type
self-consistent field (SCF) functions~\cite{23,24}. From the optical
spectra for Fe$^{3+}$ in AgX (or similar tetrahedral
environments)~\cite{25,26}, the cubic field parameters $Dq$ and the
covalency factors $N$ are obtained and shown in table~\ref{tab:1}.
\begin{table}[ht]
\caption{The group overlap integrals, the cubic field parameter
(in {cm}$^{-1}$) and the covalency factor, the normalization factors and
the orbital admixture coefficients as well as the spin-orbit coupling
coefficients (in {cm}$^{-1}$) and the orbital reduction factors in the
crystal-field and charge transfer mechanisms for Fe$^{3+}$ in AgX ({X=Cl} and {Br}).\label{tab:1}}
\vspace{2ex}
\begin{center}
{\small
\tabcolsep=0.7em
\begin{tabular}{cc cc cc cc cc cc cc cc cc }
\hline\hline
     & {Hosts} & $S_{\pi}$ & $S_{\sigma}$ & $S_{S}$ & $A$ & $Dq$
     & $N$ &  $N_{t}^\mathrm{a}$ & $N_{e}^\mathrm{a}$ & $\lambda_{t}^\mathrm{a}$
     & $\lambda_{e}^\mathrm{a}$ & $\lambda_{{s}}^\mathrm{a}$ \\
     \hline
     & {AgCl}     & $0.0069$   & $-0.0298$  & $0.0210$   & $1.489$   & $-430$   & $0.74$  & $0.729$  & $0.745$  & $0.456$  & $-0.345$  & $-0.321$ \\
     & {AgBr}     & $0.0072$   & $-0.0314$  & $0.0264$   & $1.467$   & $-380$   & $0.70$  & $0.683$  & $0.706$  & $0.516$  & $-0.380$  & $-0.312$ \\
     \hline\hline
     & {Hosts}    & $N_{t}^\mathrm{b}$ & $N_{e}^\mathrm{b}$ & $\lambda_{t}^\mathrm{b}$
     & $\lambda_{e}^\mathrm{b}$ & $\lambda_{{s}}^\mathrm{b}$ & $\zeta_\mathrm{CF}$ & $\zeta_\mathrm{CF}^{\prime}$ & $\zeta_\mathrm{CT}$
     & $\zeta_\mathrm{CT}^{\prime}$ & $k_\mathrm{CT}$ & $k_\mathrm{CT}^{\prime}$ \\
     \hline
     & {AgCl}     & $0.273$    & $0.403$    & $-0.733$   & $1.027$   & $-1.040$ & $208$   & $380$    & $482$    & $375$    & $0.813$   & $0.646$  \\
     & {AgBr}     & $0.272$    & $0.463$    & $-0.654$   & $1.040$   & $-1.081$ & $-599$  & $188$    & $1112$   & $713$    & $0.768$   & $0.630$  \\
     \hline\hline
\end{tabular}   }
\end{center}
\end{table}
The related molecular orbital coefficients $N_{\gamma}^{x}$ and
$\lambda_{\gamma}^{x}$ are determined from equations~(\ref{eq:5})--(\ref{eq:7}).
According to the free-ion values $ \zeta_{d}^{0}\approx588$~cm$^{-1}$~\cite{27} for Fe$^{3+}$ and $\zeta_{p}^{0}\approx 587$
and $2460$~cm$^{-1}$~\cite{28} for Cl$^{-}$ and Br$^{-}$,
the spin-orbit coupling coefficients and the orbital reduction
factors can be obtained for the crystal-field and charge transfer
mechanisms from equations~(\ref{eq:8}) and~(\ref{eq:9}). These values are also
listed in table~\ref{tab:1}. The Racah parameters in the energy denominators
of equation~(\ref{eq:1}) are determined from the relationships $B\approx
N^{2}B_{0}$ and $C\approx N^{2}C_{0}$~\cite{29} and the free-ion values
$B_{0} \approx 1322$ cm$^{-1}$ and $C_{0}\approx 4944$ cm$^{-1}$
\cite{27} for Fe$^{3+}$. From $\chi(\text{Fe}^{3+})\approx 2.4$,
$\chi (\text{Cl}^{-})\approx 2.8$ and $\chi(\text{Br}^{-})\approx
2.6$~\cite{11}, the charge transfer levels $E_\mathrm{n}$ and $E_\mathrm{a}$ are
calculated. Substituting the above values into equation~(\ref{eq:1}), the
$g$-shifts ({Cal}.$^{\mathrm{b}}$) for the Fe$^{3+}$
centers in AgX are obtained and presented in table~\ref{tab:2}. In order
to clarify the importance of the charge transfer contribution, the
results ({Cal}.$^{\mathrm{a}}$) containing merely the
crystal-field contribution are also presented in table~\ref{tab:2}.

\begin{table}[ht]
\caption{The $g$-shift $\Delta g$ and the superhyperfine parameters
(in $10^{-4}$ {cm}$^{-1}$) for Fe$^{3+}$ in
AgX.\label{tab:2}}
\begin{center}
\begin{tabular}{cc cc cc cc cc cc cc}
   \hline\hline
     & \ & \ & $\Delta g$ & $A^{\prime}$ & $B^{\prime}$ \\
     \hline\hline
     & \                & {Cal}.$^{\mathrm{a}}$   & $-0.0006$     & $3.9$       & $1.7$          \\
     & {AgCl}    & {Cal}.$^{\mathrm{b}}$   & $0.0128$      & $3.6$       & $2.8$          \\
     & \                & {Expt}.~\cite{9}          & $0.0133(4)$   & $3.3(5)$    & $2.0(5)$       \\
     \hline
     & \                & {Cal}.$^{\mathrm{a}}$   & $-0.0013$     & $21.1$      & $6.8$          \\
     & {AgBr}    & {Cal}.$^{\mathrm{b}}$   & $0.0420$      & $16.0$      & $8.5$          \\
     & \                & {Expt}.~\cite{9}          & $0.0427(50)$  & $16.2(15)$  & $7.8(15)$      \\
     \hline\hline
\end{tabular}
\end{center}
\begin{center}
\parbox[t]{13.5cm}{\footnotesize{
$^{\rm a}$ Calculations of the $g$ factor based only on the crystal-field contribution and those for the superhyperfine parameters of the previous studies~\cite{9} by fitting the unpaired spin densities.\\
$^{\rm b}$ Calculations based on the inclusion of both the crystal-field and charge transfer contributions.}}
\end{center}
\end{table}

\subsection*{2.2 Calculations for the superhyperfine parameters}

In the previous treatments of the superhyperfine parameters
\cite{30}, the unpaired spin densities $f_{{s}}$ and
$f_{\sigma}$$-$$f_{\pi}$ of the ligand $2s$ and $2p\sigma$ (or
$2p\pi$) orbitals were usually taken as adjustable parameters,
instead of being quantitatively correlated with the chemical bonding
between the impurity and ligands. In order to improve the above
treatments, the cluster approach~\cite{21} is applied to establish
the uniform expressions of these quantities. Thus, the
superhyperfine parameters can be written as:
\begin{eqnarray}
&&A^{\prime}=A_{{s}}+2(A_\mathrm{D}+A_{\sigma}-A_{\pi}),\nonumber\\
&&B^{\prime}=A_{{s}}-(A_\mathrm{D}+A_{\sigma}-A_{\pi}).
\label{eq:10}
\end{eqnarray}
Here $A_{{s}}$ is the isotropic contribution to the superhyperfine
parameters, charactering the effect of the ligand $ns$ orbital.
$A_\mathrm{D}$ and $A_{\sigma}-A_{\pi}$ denote the anisotropic
contributions from the dipole-dipole interaction between the
electron of the central ion and ligand nucleus and that from the
ligand $np$ orbital, respectively. The isotropic part can be expanded
as~\cite{30}:
\begin{eqnarray}
&&A_{{s}}=f_{{s}}A_{{s}}^{0}/(2S),
\label{eq:11}
\end{eqnarray}
with $A_{{s}}^{0} = (8/3)g_{{s}}g_\mathrm{n}\beta\beta_\mathrm{n}
|\Psi(0)|^{2}\approx 555.7\times 10^{-4}$ and 7815.0~cm$^{-1}$, and
$A_{p}^{0} = g_{{s}}g_\mathrm{n}\beta\beta_\mathrm{n}\langle r^{-3}\rangle\approx 46.7\times
10^{-4}$ and $232.2\times10^{-4}$ for Cl$^{-}$ and
Br$^{-}$, respectively~\cite{31}. $f_{{s}}$ denotes the unpaired
spin-density of the ligand $ns$ orbital. The electron
 spin is $S$ =5/2 for the ground state $^{6}{A}_{1{g}}$. The anisotropic contribution from
 the ligand $np$ orbital is usually expressed as~\cite{30}:
\begin{eqnarray}
&&A_{\sigma}-A_{\pi}=A_{p}^{0}(f_{\sigma}-f_{\pi})/(2S).
\label{eq:12}
\end{eqnarray}
Here, $f_{\sigma}$ and $f_{\pi}$ are the unpaired spin
densities of the ligand $np\sigma$ and $np\pi$ orbitals,
respectively. The dipole-dipole interaction between the electron
distribution of the central ion and the halogen ligand nucleus can
be expressed as $A_\mathrm{D} =g\beta g_\mathrm{n}\beta_\mathrm{n}/R^{3}$, with the $g$
factor of the central ion. In the above expressions, the ligand
unpaired spin densities can be quantitatively connected with the
relevant molecular orbital coefficients based on the cluster
approach:
\begin{eqnarray}
&&f_{{s}}\approx N_{e}^\mathrm{a}(\lambda_{{s}}^\mathrm{a})^{2}/3\,, \qquad
f_{\sigma}\approx N_{e}^\mathrm{a}(\lambda_{e}^\mathrm{a})^{2}/3\,, \qquad
f_{\pi}\approx N_{t}^\mathrm{a}(\lambda_{t}^\mathrm{a})^{2}/4\,.
\label{eq:13}
\end{eqnarray}

It is noted that in the previous works~\cite{9,30} the unpaired
spin densities were simply treated as the adjustable parameters by
fitting the experimental superhyperfine parameters. Instead, they
are quantitatively and uniformly calculated from the cluster
approach in this work.

The unpaired spin densities $f_{i}$ $(i =\sigma, \pi, s)$ as
well as the isotropic contribution $A_{{s}}$ and the anisotropic
contribution $A_{\sigma}- A_{\pi}$ and $A_{\mathrm{D}}$ to the
superhyperfine parameters are acquired from  equations~(\ref{eq:11})--(\ref{eq:13}), and
thus the resultant $A^{\prime}$ and $B^{\prime}$ are obtained from equation~(\ref{eq:10}).
In addition, by fitting the unpaired spin densities, the theoretical results of the previous work~\cite{9}  are also collected in table~\ref{tab:2}.

\section*{3. Discussion}

Table~\ref{tab:2} reveals that the theoretical $g$ factors
({Cal}.$^{ \mathrm{b}}$) for AgX:Fe$^{3+}$ based
on the inclusion of both the crystal-field and charge transfer
contributions show reasonable agreement with the experimental data,
whereas those ({Cal}.$^{ \mathrm{a}}$) based only on the
conventional crystal-field contribution are merely $3\%-5\%$ of
the observed values. Meanwhile, the superhyperfine parameters are
also suitably analyzed from the uniform quantitative relationships
between the unpaired spin densities and the relevant molecular
orbital coefficients based on the cluster approach.

1) The charge transfer contribution to the $g$-shift has
opposite (positive) sign and much larger magnitude as compared to
the crystal-field one. With an increase of the ligand spin-orbit
coupling coefficient $ \zeta_{p}^{0}$ and a decease of the covalency
factor $N$, the importance of the charge transfer contribution
(characterized by the relative ratio  $\Delta g_\mathrm{CT}/\Delta g_\mathrm{CF}$)
increases rapidly from 13 for Cl$^{-}$ to 34 for
Br$^{-}$. So, for $3d^5$ ions (especially
Fe$^{3+}$ with high valence state) in covalent hosts, the
charge transfer contribution to the $g$-shift is significant due to
the low charge transfer levels $E_\mathrm{n}$ and $E_\mathrm{a}$.

2) Apart from the increase of the charge transfer contribution
from Cl$^{-}$ to Br$^{-}$, the  $\Delta g_\mathrm{CF}$ from
the crystal-field contribution also shows a similar but less
significant tendency (see table~\ref{tab:2}). This can be ascribed to the
relative difference (or anisotropy) between $\zeta_\mathrm{CF}$ and
$\zeta_\mathrm{CF}^{\prime}$ in the formula of $\Delta g_\mathrm{CF}$, which is
relevant to the low covalency factors $N$ ($\approx 0.7\ll 1$,
which are much smaller than the values 0.90 and 0.87 for
Fe$^{3+}$ in fluorides and oxides~\cite{32}) and the obvious
orbital admixture coefficients ($\approx 0.3-0.5$) as well as the
large $\zeta_{p}^{0}$. Especially, the ratios
$\zeta_\mathrm{CF}/\zeta_\mathrm{CF}^{\prime}$ are $83\%$ and $165\%$ for {AgCl}
and {AgBr}, increasing rapidly with the increase of
$\zeta_{p}^{0}$. Therefore, the impurity-ligand orbital admixtures
and the anisotropic contribution from the ligand spin-orbit coupling
coefficient should be taken into account in the analysis of the $g$
factors for AgX:Fe$^{3+}$. Further, the relatively
weaker dependence of  $\Delta g_\mathrm{CF}$ on the covalency or the ligand
contributions than $\Delta g_\mathrm{CT}$ is attributable to the dominant
third-order perturbation terms (inversely proportional to the square
of the energy separation $E_{1}$, $E_{2}$ or $E_{3}$) in the former
and the second-order perturbation terms (inversely proportional to
the charge transfer levels $E_\mathrm{n}    $ or $E_\mathrm{a}$) in the latter.

3) Reasonable agreement between theory
({Cal}.$^{\mathrm{b}}$) and experiment is achieved for $A^{\prime}$,
but slightly worse for $B^{\prime}$. This may be ascribed to the errors
arising from the theoretical model (e.g., the ligand field model and
the cluster approach) and formulas. Importantly, the present
calculations are based on the uniform model and formulas by
establishing the quantitative relationships between the unpaired
spin densities and the relevant molecular orbital coefficients from
the cluster approach. Thus, the quantities $f_{{s}}$ and $f_{\sigma}$
$-$ $f_{\pi}$ in equations~(\ref{eq:11}) and (\ref{eq:12}) are determined
theoretically from the cluster approach in a uniform way. These quantities were normally taken as the adjustable parameters by
fitting two experimental superhyperfine parameters in the previous
works~\cite{9,30}. The calculated values of $f_{{s}}$ ($\approx 0.6 \%$ and
$0.7\%$) and $f_{\sigma}-f_{\pi}$ ($\approx 1.95\%$ and
$1.6\%$) for {AgCl}:Fe$^{3+}$ and
{AgBr}:Fe$^{3+}$ in this work are comparable with the
estimated values ($f_{{s}}\approx 0.86\%$ and $0.74\%$, and
$f_{\sigma}-f_{\pi}\approx 3.4\%$ and $6.5\%$,
respectively) by directly fitting two experimental superhyperfine
parameters in the previous study~\cite{9}, while the lower unpaired spin
densities yield slightly better results. Apparently, the
theoretical model and formulas in this work can also be applied to
the EPR analysis for $3d^5$ ions in other fluorides.

4) There are some errors in the present calculations, e.g.,
the theoretical $B^{\prime}$ is slightly larger than the observed value
in view of the experimental uncertainty. The errors may be ascribed
to the approximations of the theoretical model (e.g., the ligand
field theory and the cluster approach) and formulas, i.e., only the
central ion $3{d}$ orbitals and the valence ($ns$ and $np$, with $n
= 3$ or 4 for Cl$^{-}$ or Br$^{-}$) orbitals of the
nearest neighbour ligands are included in the cluster approach
calculations. All the intermediate parameters ($\zeta_\mathrm{CF}$,
$\zeta_\mathrm{CF}^{\prime}$, $\zeta_\mathrm{CT}$, $\zeta_\mathrm{CT}^{\prime}$, $k_\mathrm{CT}$,
$k_\mathrm{CT}^{\prime}$) are quantitatively determined from the related cluster
approach formulas. Except the spectral parameters $Dq$ and $N$
obtained from the optical spectral measurements, no adjustable
parameters are induced in the calculations. Of course, the
theoretical calculations and results in this work should be regarded
as tentative ones. In order to make further investigations on the
EPR spectra (especially the superhyperfine parameters) for
{AgX:Fe}$^{3+}$, one may adopt more powerful and reliable
density function theory (DFT) treatments~\cite{33,34,35}.

\section*{4. Summary}

The EPR parameters of {AgX:Fe}$^{3+}$ are theoretically
studied from the perturbation formulas containing both the
crystal-field and charge transfer contributions. The $g$-shift
$\Delta g$ from the charge transfer contribution is opposite
(positive) in sign and much larger in magnitude as compared to
that from the crystal-field one. Moreover, the importance of the
charge transfer contribution increases rapidly with an increase of the
covalency and the spin-orbit coupling coefficient of the ligand,
i.e., $\text{Cl}^{-} < \text{Br}^{-}$. The unpaired spin densities
are quantitatively obtained from the relevant molecular orbital
coefficients using the cluster approach instead of being treated as
adjustable parameters by fitting the experimental superhyperfine
parameters in the previous works.

\section*{Acknowledgements}

This work was financially supported by ``the Fundamental
Research Funds for the Central Universities'' under granted No.
ZYGX2010J047 and No. E022050205.

\ukrainianpart

\title{Дослідження параметрів  EPR для тетраедральних кластерів
[FeX$_4$]$^{-}$ у  AgX (X${}={}$Cl, Br)}

\author{Б.-T. Сонг\refaddr{ad1}, С.-Й.~Ву\refaddr{ad1,ad2},
M.-К.~Куанг\refaddr{ad1}, З.-Х.~Жанг\refaddr{ad1}}

\addresses{\addr{ad1} Кафедра прикладної фізики, Китайський університет електроніки та технологій,  Ченду 610054, КНР
\addr{ad2} Міжнародний центр фізики матеріалів, Академія наук Китаю, Шеньян 110016, КНР}

\newpage

\makeukrtitle

\begin{abstract}
\tolerance=3000%
Параметри  $g$ факторів електронного парамагнітного резонансу і супертонкі параметри для
тетраедральних  кластерів [FeX$_4$]$^{-}$ у  AgX (X${}={}$Cl, Br)
досліджуються теоретично, використовуючи  теорію збурень для цих параметрів
для  $3d^5$ іона в тетраедрі з врахуванням вкладів від
кристалічного поля і перенесення заряду. Параметри моделі
кількісно визначаються з кластерного підходу в єдиний спосіб.
Вклад у $g$-зсув ${\triangle}g$ ($=g-g_{\mathrm{s}}\,$, де
$g_{\mathrm{s}}\approx 2.0023$~--- значення спіну), отриманий із врахування
зарядового перенесення є протилежний (позитивний) за знаком і
набагато більший  за величиною в порівнянні з вкладом, отриманим від врахування
кристалічного поля. Важливість врахування зарядового перенесення
зростає з ростом ковалентності і коефіцієнту
спін-орбітальної взаємодії ліганди і отже демонструє, що
AgCl${}<{}$AgBr. Густини неспарених спінів галогенів, $np\sigma$  і
$np\pi$  орбіталей кількісно визначаються з молекулярних
орбітальних коефіцієнтів, що базуються на кластерному підході.

\keywords гамільтоніани кристалічних полів і спінів,
електронний парамагнетний резонанс  (EPR)

\end{abstract}

\end{document}